# HISTORICAL TRAFFIC FLOW DATA RECONSTRUCION APPLYING WAVELET TRANSFORM

# RECONSTITUIÇÃO DE DADOS HISTÓRICOS DE FLUXO DE TRÁFEGO UTILIZANDO TRANSFORMADA *WAVELET*

**Elaine Rodrigues Ribeiro**
**André Luiz Cunha**
Universidade de São Paulo
Escola de Engenharia de São Carlos

**ABSTRACT**
Despite the importance of fundamental parameters (traffic flow, density and speed) to describe the traffic behavior, there still are some difficulties in order to obtain and store this information. Furthermore, given the type of study or the project the resolution analysis interval can vary from less than one hour to annual. To create alternatives in database structures, this article aims to present a method to reconstruct disaggregated historical data from aggregated data using Wavelet Transform. From the proposed method, it is possible to reconstruct data in short intervals from data with longer intervals, since they have the same behavior, for example, data from the same or similar highway. For such, a Detail coefficient is generated through the Discrete Wavelet Transform (DWT) with the disaggregated data. The aggregated data was reconstructed through an Approximation coefficient. After establishing these coefficients, the Inverse Wavelet Transform (IWT) is applied. The results indicated an average correlation between the reconstructed the original data of 0.960; 0.974; 0.968 and 0.960 for data initially aggregated at 10, 20, 40 and 80 minutes interval, respectively. The results also indicated an absolute mean error of 7.13%; 8.74%; 9.83% and 11.23% for data initially aggregated in 10, 20, 40 and 80 minutes, respectively. In other words, results suggest that the reconstituted data have a high correlation and a low percentage of mean absolute error with the original signal. In conclusion, the reconstruction of data from aggregated data and Wavelet Transform presents a good correlation and low average absolute error rate, comparable to traffic estimation studies (CASTRO-NETO et al., 2009; CORIC et al. 2012, LAM et al., 2006; LIM, 2001).

**RESUMO**
Apesar da importância de parâmetros básicos (fluxo de tráfego, densidade e velocidade) em descrever o comportamento do tráfego, ainda há dificuldades para obter e armazenar essas informações. Além disso, dado o tipo de estudo ou projeto, a resolução do intervalo de análise pode variar de menos de uma hora a anual. Considerando isso e com o intuito de criar alternativas para estruturar um banco de dados, o objetivo deste artigo é apresentar um método para reconstruir dados históricos desagregados a partir de dados agregados utilizando Transformada *Wavelet*. A partir do método proposto é possível reconstruir dados em intervalos curtos a partir de dados com intervalos maiores, desde que possuam o mesmo comportamento, por exemplo, dados de uma mesma rodovia ou similar. Para tal um coeficiente de Detalhe é gerado com os dados agregados por meio da Transformada *Wavelet* Discreta (TWD, do inglês *Discrete Wavelet Transform* – DWT). Os dados agregados obtidos do dia que se deseja reconstruir são utilizados como coeficiente de Aproximação. Após estabelecer esses dois coeficientes, a Transformada *Wavelet* Inversa (TWI, do inglês *Inverse Wavelet Transform* – IWT) foi aplicada. Os resultados indicaram uma correlação média entre os dados reconstruídos e os dados originais de 0,960; 0,974; 0,968 e 0,960 para dados agregados inicialmente em intervalos de 10, 20, 40 e 80 minutos, respectivamente. Também foi identificado um erro médio absoluto de 7,13%; 8,74%; 9,83% e 11,23% para os dados agregados incialmente em 10, 20, 40 e 80 minutos, respectivamente. Ou seja, esses resultados sugerem que (*i*) os dados reconstituídos têm uma alta correlação e (*ii*) uma baixa porcentagem de erro médio absoluto com o sinal original. Conclui-se que a reconstrução de dados a partir de dados agregados e Transformada *Wavelet* apresenta uma boa correlação e baixa taxa de erro médio absoluto, comparável a estudos de estimativa de tráfego (CASTRO-NETO et al., 2009; CORIC et al. 2012, LAM et al., 2006; LIM, 2001).

## 1. INTRODUÇÃO
Além dos problemas relacionados à obtenção de dados, a resolução exigida a cada parâmetro analisado é distinta. Embora as variações no volume de tráfego possam ser analisadas em várias escalas (minuto, hora, dia, mês e ano), para a análise do comportamento do tráfego é comum utilizar valores representativos da corrente de tráfego, tais como Volume Diário (em veic/dia) e Volume Horário (em veic/h) (MAY, 1990). No Brasil, o Manual de Estudos de Tráfego (DNIT, 2006) exemplifica as variações do volume em diferentes escalas de tempo e apresenta conceitos de Volume Diário Médio Anual (VDMA), Volume Diário Médio (VDM) e Volume Horário Médio (VHM). O uso desses parâmetros médios e suas derivações pode ser insuficiente na análise comportamental do tráfego, pois podem suavizar ou distorcer anomalias no fluxo de tráfego, perdendo assim informações importantes para a definição de dias atípicos (RIBEIRO; CUNHA, 2016). Em análises de previsão de tráfego a curto prazo, por exemplo, a resolução (nível de agregação) interfere diretamente nos modelos de previsão (DOUGHERTY; COBETT, 1997): dados com baixa resolução fornecem menos informações sobre o tráfego, enquanto dados de alta resolução tendem a ser ruidosos, gerando erros de previsão (ABDULHAI *et al.,* 2002).

Uma alternativa para reduzir ou até mesmo eliminar as limitações que as diferentes resoluções podem ocasionar é o processamento digital de sinais, em específico a Transformada *Wavelet*. As *Wavelets* são capazes de decompor um sinal no domínio do tempo em uma série de dados em diferentes escalas de frequência e de tempo. No entanto, por ser uma técnica recentemente empregada na prática da Engenharia de Transportes, ainda faltam fundamentos teóricos e orientações do seu uso (ZHENG; WHASHINGTON, 2012).

Dada a influência e importância que os dados históricos do fluxo de tráfego têm para diversas linhas de estudo, pesquisa e projeto para Engenharia de Transportes, o presente trabalho apresenta um método de reconstituição de dados de tráfego desagregados a partir de dados agregados utilizando Transformada Wavelets. Busca-se apresentar uma alternativa para análise de parâmetros que descrevam o comportamento do tráfego mesmo quando os dados históricos contenham falhas ou baixa frequência de aquisição. Para tanto, este artigo está estruturado da seguinte forma: a Seção 2 apresenta a fundamentação teórica da pesquisa com uma revisão sobre a Transformada *Wavelet* e trabalhos aplicados em Engenharia de Transportes que utilizaram a mesma; a Seção 3 descreve os dados utilizados para aplicação do método; a Seção 4 aborda o método utilizado para reconstituição de informações do fluxo de tráfego em baixo nível de agregação; a Seção 5 discute as análises e resultados obtidos bem como suas limitações e, por fim, a Seção 6 conclui o artigo com recomendações e perspectivas de trabalhos futuros que tenham como objetivo o aprimoramento das análises e tratamento dos bancos dados disponíveis.

## 2. REVISÃO BIBLIOGRÁFICA
### 2.1. Resolução dos parâmetros de tráfego
Em Engenharia de Transportes a etapa de obtenção e análise de dados é fundamental pois estudos e projetos são altamente dependentes da aquisição de informações do sistema. Esses dados podem ser obtidos de duas formas, manual ou automática. No levantamento automático, esses dados são coletados continuamente ao longo do tempo (dados contínuos) e então são convertidos por amostragem em categorias em função do tempo (dados discretos). Portanto, um banco de dados contendo parâmetros de tráfego amostrados em intervalos determinados de tempo é um sinal discreto. Assim sendo, a resolução dos dados (nível de agregação do conjunto de dados) analisados pode ser determinada conforme sua finalidade.

Os estudos dos parâmetros do tráfego podem ser em período anual, mensal, diário, horário e até subhorário. Além dos conceitos de VDMA, VDM e VHM para o estudo de volume de tráfego, o *Highway Capacity Manual* (HCM) (TRB, 2019) também apresenta a definição de taxa de fluxo de tráfego e o Fator Hora Pico (FHP, do inglês *Peak Hour Fator* – PHF). A taxa de fluxo de tráfego descreve o número de veículos observados em uma seção durante um intervalo de tempo menor que uma hora, normalmente em 15 minutos, mas expresso em taxa horária equivalente. O conceito de FHP expressa a flutuação dentro da hora de pico, sendo calculado pela relação entre o Volume Horário (VH) na hora pico pela taxa de fluxo equivalente nos 15 minutos mais congestionado da hora pico. No entanto, a flutuação do fluxo de tráfego pode ocorrer em outras resoluções, além da variação de 15 minutos indicada pelo HCM (TRB, 2019).

As características inerentes ao fluxo de tráfego são a sua variação temporal – minuto, hora, dia, semana, mês e ano – e espacial – em diferentes locais, sentido de tráfego e segundo a faixa de tráfego analisada. Essas variações podem ocorrer por efeitos conhecidos e predeterminados, como por exemplo o horário de pico em um dia, feriados, ou segundo as características geométricas e funcionais da via. O outro conjunto de fatores não predeterminadas, mas as causas são conhecidas, são as condições de tempo (CARDOSO *et al.,* 2019) ou a ocorrência de incidentes (OLIVEIRA, 2004).

É importante destacar que assim que os dados são discretizados em determinado intervalo de tempo podem ocorrer um ou mais desses três problemas: (1) a suavização ou distorção das características do fluxo de tráfego dentro da agregação estabelecida, perdendo informações importantes para a definição do comportamento de tráfego e dias atípicos (ZHENG *et al.,* 2011; ZHENG; WASHINGTON, 2012); (2) criar uma base de dados ruidosa que diminui a capacidade de previsão dos modelos de previsão baseados em dados históricos (CASTRO-NETO *et al.,* 2009); e (3) a impossibilidade de desagregar uma informação agregada, muitas vezes representada apenas pela média.

Outra forma de analisar os parâmetros de tráfego é a partir do processamento digital de sinais. Tal processamento, pode ocorrer tanto no domínio da frequência quanto no domínio do tempo. No domínio da frequência o sinal é decomposto em relação as frequências e é analisada a amplitude de cada uma separadamente sem a informação do instante em que ocorrem. No domínio do tempo, o sinal é mantido e é estuda a amplitude do sinal a cada instante de tempo. A Figura 1 ilustra a diferença entre os dois domínios para o mesmo sinal.

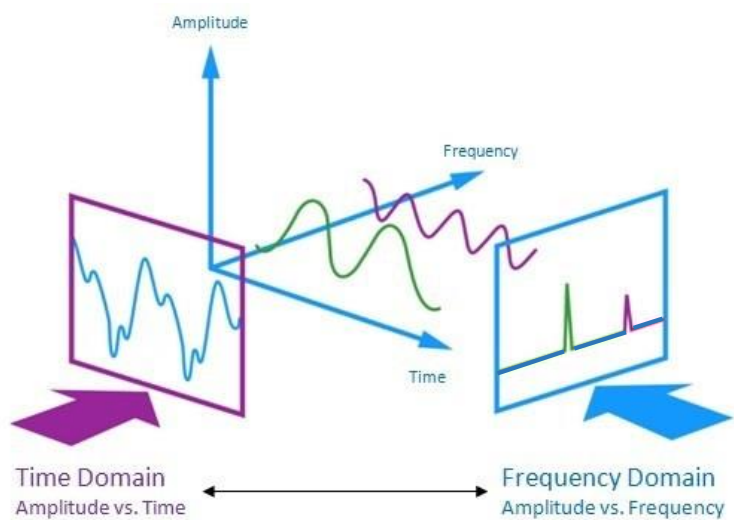

**Figura 1:** Sinal no domínio do tempo e no domínio da frequência (MED, 2018)

### 2.2. Transformada *Wavelet*

Para analisar sinais estacionários (sinal que possui sempre a mesma frequência ao longo de todo o período de tempo) é comum utilizar a Transformada de Fourier. Essa técnica contém o maior número de aplicações práticas, utilizada nas mais diversas áreas de ciências aplicadas, sendo fundamental em telecomunicações e no processamento digital de imagens (BARBOSA; BLITZKOW, 2008). A vantagem da Transformada Discreta de Fourier (TDF, do inglês *Discrete Fourier Transform* – DFT) é permitir, a partir de um sinal no domínio do tempo, obter o sinal correspondente no domínio da frequência. Entretanto, a desvantagem da TDF é que nada se conhece da localização dessas frequências, o que pode ser um problema quando analisamos um sinal não estacionário (GUIDO *et al.,* 2012). Apesar da evolução dessa técnica para a Transformada Janelada de Fourier (TJF), aplicação de análise de Fourier deslocando uma janela móvel no tempo, ainda assim existem limitações (BARBOSA; BLILTZKOW, 2008). Portanto, a técnica denominada *Wavelets* se desenvolveu da necessidade de se conhecer onde as frequências ocorrem.

As *Wavelets* são funções capazes de decompor e descrever uma série de dados no domínio do tempo em diferentes escalas de frequência e de tempo. Para que uma função seja denominada de função *Wavelet*, ela deve atender a duas propriedades distintas:

a) A integral da *Mother Wavelet* deve ser igual a zero;

$$\int_{-\infty}^{+\infty} \psi(t)dt = 0 \quad (1)$$

b) A função *Wavelet* deve possuir energia unitária;

$$\int_{-\infty}^{+\infty} |\psi(t)|^2 = 1 \quad (2)$$

As propriedades citadas acima expressam que $\psi(t)$ é quadraticamente integrável, ou seja, a função *Mother Wavelet* pertence ao conjunto das funções de quadrado integrável $L^2(\mathbb{R})$ dentro do conjunto de números reais $\mathbb{R}$. As propriedades acima, sugerem também que $\psi(t)$ tende a oscilar acima e abaixo do eixo *t* e que tem sua energia localizada em uma certa região, pois ela é finita. A maior parte de sinais reais de importância na aplicação de processamento de sinais são encontrados nessa classe de funções finitas (BARBOSA; BLITZKOW, 2008).

A aplicação das funções *Wavelets* inclui a Transformada *Wavelet* Contínua (TWC, do inglês *Continuous Wavelet Transform* – CWT) e a Transformada *Wavelet* Discreta (TWD, do inglês *Discrete Wavelet Transform* – DWT), além de suas respectivas Transformadas Inversas. A TWC traz muitas repetições de informações do sinal analisado, tornando-a desinteressante computacionalmente, no entanto a TWD é utilizada na chamada "análise de multiresolução" ou em uma versão mais completa que permite um detalhamento personalizado do espectro, chamada de "análise por pacotes". Uma das estruturas de Transformação Discreta mais utilizadas entre as disponíveis na literatura é as "diádicas", que operam com escalas e posições baseadas em potências de 2 (dois) (DIAS, 2003). A TWD decompõe o sinal analisado em componentes de alta escala, nomeados de "Aproximações" e de baixa escala, nomeados de "Detalhes", em cada resolução. O componente de Aproximação, corresponde ao conteúdo de baixa frequência do sinal, gerado pelo filtro passa-baixa e o componente de Detalhe corresponde ao conteúdo de alta frequência do sinal, gerado pelo filtro passa-alta (GUIDO *et al.,* 2012).

Se cada sinal analisado gerasse duas novas componentes de tamanho igual ao sinal original, o resultado de um DWT apresentaria, portanto, duas vezes mais dados do que o sinal original e assim sucessivamente a cada nível. No entanto, a decomposição é realizada reduzindo a taxa de amostragem (*downsampling*) na mesma proporção que o número de decomposições (DIAS, 2003).

A aplicação da TWD em *k* níveis resulta sempre em um sinal de tamanho *N*, composto de uma aproximação no nível k ($A_k$) e k detalhes ($D_k, D_{k-1}, ..., D_1$). A decomposição no nível 1, resulta nos componentes $A_1$ e $D_1$; no nível 2, o resultado é composto pelos componentes $A_2, D_2$ e $D_1$; no nível 3, os componentes são definidos por $A_3, D_3, D_2$ e $D_1$. A estrutura hierárquica da TWD garante a sua propriedade de decomposição e reconstrução do sinal *S*.

$$d_{j,k} = \langle \psi_{j,k}(t), f(t) \rangle = \int f(t)\psi_{j,k}(t)dt \quad (3)$$

$$f(t) = \sum_{j}^{\infty} = -\infty \sum_{k}^{\infty} = -\infty \psi_{j,k}(t)d_{j,k} \quad (4)$$

Sendo d*j,k* os Componentes da *Wavelet*, correspondentes ao sinal transformado F(*a,b*) em que *a* é a escala e *b* o deslocamento (FARIA, 1997).

Nos últimos anos houve grande desenvolvimento matemático das *Wavelets* e suas aplicações para diversas áreas. Devido à variação temporal do comportamento do tráfego, a Transforma *Wavelet* se tornou uma ferramenta eficaz de decomposição tempo-frequência (ZHENG; WHASHINGTON,

2012). Na área de Engenharia de Transportes a aplicação das *Wavelets* combinadas com técnicas de mineração de dados – tais como, análise de agrupamento, lógica fuzzy, redes neurais – tem se concentrado em questões relacionadas a: (a) Detecção automática de incidentes em autoestrada (ADELI; SAMANT, 2000; BELLO-SALAU *et al.,* 2018; GHOSH-DASTIDAR; ADELI, 2003; JEONG *et al.,* 2009; KARIM; ADELI, 2002; KARIM; ADELI, 2003; SAMANT; ADELI, 2000; SAMANT; ADELI, 2001; TENG; QI, 2003); (b) Característica do tráfego em torno de autoestrada em zonas de trabalho (ADELI; GHOSH-DASTIDAR, 2004; GHOSH-DASTIDAR; ADELI, 2006; GHOSH *et al.,* 2010); (c) Previsão do fluxo de tráfego (BOTO-GIRALDA *et al.,* 2010, HOU *et al.,* 2019; JIANG; ADELI, 2005; PENG; XIANG, 2019; XIE *et al.,* 2007; ZHANG, 2016); (d) Reconhecimento de padrões de tráfego (CHUNG, 2003; GHOSH *et al.,* 2006; RABBOUCH *et al.,* 2018; RIBEIRO; CUNHA 2016; RIBEIRO, 2017); e (e) Detecção de oscilações do tráfego (MOHAN *et al.,* 2014; ZHENG *et al.,* 2011; ZHENG; WHASHINGTON, 2012; RIBEIRO; CUNHA, 2017; RIBEIRO, 2017).

**3. BANCO DE DADOS**
Os dados utilizados neste artigo foram obtidos por sensores de laço indutivo que coligiam informações de fluxo e velocidade média da corrente de tráfego, entre o período de setembro/2011 a maio/2014. Por intermédio da ARTESP (Agência Reguladora de Serviços Públicos de Transporte do Estado de São Paulo), o banco de dados foi fornecido pelo grupo CCR de concessionárias de rodovias do estado de São Paulo, sendo: a CCR Via Oeste responsável pelas Rodovia Presidente Castelo Branco (SP-280), Rodovia Raposo Tavares (SP-270); CCR AutoBan opera a Rodovia dos Bandeirantes (SP-348) e CCR Rodoanel responsável pelo anel viário Mário Covas (SP-021).

Embora as quatro rodovias sejam operadas pelo mesmo grupo, cada concessionária captura e registra os dados de tráfego de forma diferente. Os dados disponibilizados pela concessionária CCR Via Oeste e CCR Rodoanel era composto por planilhas com informações de data, hora, volume total, volume de automóveis, volume de comerciais e volume de motos, assim como as respectivas velocidades média dessas categorias veiculares. Por outro lado, os dados disponibilizados pela CCR AutoBan eram planilhas contendo a data e hora, volume total, volume de automóveis e veículos comerciais, taxa de ocupação e velocidade média de todos os veículos. Enquanto a CCR Via Oeste forneceu dados em planilhas bimestrais agregados em cinco minutos, a concessionária CCR AutoBan forneceu dados em planilhas mensais agregados em seis minutos.

Após a obtenção dos dados, o primeiro passo foi padronizar o banco de dados eliminando dados inconsistentes (seja com ausência de informação ou dados de tráfego repetidos erroneamente) que poderiam interferir na descrição das condições de tráfego local. Foram utilizados 20 sensores contendo as informações de tráfego de ambos sentidos da via, totalizando 685 planilhas organizadas e filtradas. A Tabela 1 apresenta os sensores analisados e a quantidade de dados faltantes mensalmente para o período de um ano com a menor quantidade de dados faltantes (setembro/2011 a agosto/2012).

Ao analisar a Tabela 1, é possível notar a quantidade de dados faltantes mesmo para o ano com a menor quantidade de falhas. Apesar do extenso banco de dados, nenhum sensor completou um ano inteiro sem falhas. O pior sensor (Rodoanel Mário Covas, km 22,3 ExPr) não coletou nenhum dado para o mesmo período. Isso demonstra a dificuldade inerente de coletas de dados constante, sem falhas, em longos períodos e em pontos de coleta distintos. Para aplicação da Transformada *Wavelet* são necessárias séries temporais completa, sem falhas. Para tal, foi definido o sensor localizado na Rodovia Presidente Castelo Branco (SP-280) no km 51,9 e os poucos dados faltantes foram substituídos por zero.

**Tabela 1:** Quantidade de dados faltantes mensalmente por sensor no período de um ano

| RODOVIA | KM | FAIXA | set/11 | out/11 | nov/11 | dez/11 | jan/12 | fev/12 | mar/12 | abr/12 | mai/12 | jun/12 | jul/12 | ago/12 |
|---|---|---|---|---|---|---|---|---|---|---|---|---|---|---|
| Presidente Castelo Branco | 15,9 | OM | 0 | 0 | 0 | 0 | 194 | 12 | 11 | 323 | 445 | 3.306 | 0 | 0 |
| | 16,0 | L/OE | 0 | 12 | 0 | 0 | 65 | 828 | 0 | 23 | 0 | 41 | 0 | 0 |
| | | LM | 293 | 1.830 | 0 | 0 | 0 | 1.429 | 333 | 201 | 0 | 2.221 | 0 | 0 |
| | 18,4 | LM | 0 | 25 | 0 | 430 | 528 | 830 | 0 | 0 | 0 | 0 | 12 | 0 |
| | 22,3 | L/OE | 0 | 12 | 0 | 0 | 39 | 1.116 | 1.567 | 384 | 618 | 0 | 0 | 0 |
| | | LM | 0 | 12 | 0 | 0 | 0 | 1.150 | 0 | 0 | 0 | 0 | 0 | 0 |
| | | OM | 0 | 12 | 0 | 0 | 0 | 1.176 | 788 | 0 | 0 | 130 | 1.151 | 130 |
| | 26,9 | L/OE | 1.095 | 8.928 | 8.640 | 8.928 | 8.928 | 8.352 | 8.928 | 8.640 | 933 | 3.160 | 0 | 40 |
| | 29,5 | L/OE | 0 | 12 | 0 | 0 | 33 | 12 | 0 | 0 | 0 | 0 | 0 | 0 |
| | 37,0 | L/OE | 3 | 12 | 26 | 35 | 0 | 12 | 59 | 314 | 39 | 139 | 0 | 0 |
| | 51,9 | L/OE | 0 | 12 | 0 | 0 | 0 | 12 | 0 | 0 | 0 | 0 | 0 | 0 |
| | 59,6 | L/OE | 0 | 295 | 0 | 0 | 18 | 1.182 | 1.099 | 4.707 | 5.451 | 7.854 | 4.069 | 372 |
| | 75,0 | L/OE | 0 | 1.177 | 2.660 | 0 | 0 | 1.421 | 0 | 282 | 5 | 2.259 | 5.582 | 3.056 |
| Raposo Tavares | 34,0 | L/OE | 8.640 | 8.928 | 8.640 | 8.928 | 1.248 | 80 | 39 | 0 | 0 | 0 | 0 | 0 |
| | 39,9 | L/OE | 0 | 22 | 10 | 908 | 1.731 | 423 | 94 | 42 | 30 | 0 | 0 | 75 |
| Rodoanel Mário Covas | 18,3 | Ex/InPr | 8.640 | 2.217 | 1.722 | 2.808 | 258 | 0 | 2.182 | 250 | 949 | 40 | 0 | 26 |
| | 22,3 | ExPr | 8.640 | 8.928 | 8.640 | 8.928 | 8.928 | 8.352 | 8.928 | 8.640 | 8.928 | 8.640 | 8.928 | 8.928 |
| | | InPr | 42 | 13 | 0 | 226 | 1.101 | 15 | 30 | 1.872 | 8.928 | 3.393 | 1.447 | 838 |
| AutoBan | 32,0 | N | 1.700 | 110 | 1.660 | 320 | 7.440 | 103 | 37 | 1.220 | 51 | 50 | 70 | 50 |
| | | S | 0 | 1.090 | 1.660 | 320 | 7.440 | 103 | 37 | 90 | 51 | 50 | 70 | 46 |
| | 50,0 | N | 3.890 | 10 | 7.200 | 4.420 | 890 | 312 | 105 | 470 | 590 | 780 | 0 | 10 |
| | | S | 0 | 10 | 7.200 | 3.020 | 926 | 312 | 105 | 470 | 590 | 780 | 0 | 10 |
| | 65,0 | N | 0 | 10 | 2.530 | 4.500 | 0 | 10 | 567 | 2.140 | 0 | 0 | 280 | 120 |
| | | S | 5.200 | 600 | 2.530 | 300 | 0 | 10 | 567 | 280 | 0 | 0 | 280 | 120 |
| | 92,0 | N | 2.830 | 830 | 730 | 4.810 | 30 | 15 | 6.010 | 7.200 | 2.210 | 0 | 36 | 0 |
| | | S | 2.830 | 830 | 730 | 2.970 | 30 | 15 | 0 | 170 | 0 | 0 | 36 | 0 |
| | 131,0 | N | 0 | 300 | 1.581 | 70 | 1.616 | 14 | 0 | 60 | 0 | 0 | 0 | 0 |
| | | S | 0 | 300 | 1.581 | 1.180 | 1.616 | 14 | 0 | 60 | 0 | 0 | 0 | 0 |
| | 149,0 | N | 1.840 | 10 | 0 | 480 | 410 | 631 | 522 | 10 | 0 | 0 | 0 | 380 |
| | | S | 1.840 | 10 | 0 | 480 | 410 | 631 | 522 | 10 | 0 | 0 | 0 | 380 |

**Legenda:** N (Norte), S (Sul), L (Leste), O (Oeste), M (Marginal), E (Expressa), In (Interna), Ex (Externa) e P (Principal)

**Legenda de Cores:**
- 🟢 Até uma hora de dados faltantes
- 🟡 De uma hora a um dia de dados faltantes
- 🟠 De um dia a uma semana de dados faltantes
- 🔴 Mais de uma semana de dados faltantes

## 4. MÉTODO PROPOSTO

O objetivo deste trabalho é reconstruir um sinal desagregado (5 minutos) a partir de dados agregados (80 minutos) utilizando Transformada *Wavelet*. Para aplicação da Transformada *Wavelet* Inversa (TWI) são necessários dois componentes conhecidos, Aproximação e Detalhe. No método proposto, o coeficiente de Detalhe é gerado a partir da Transformada *Wavelet* Discreta (TWD) e o coeficiente de Aproximação é o sinal desagregado do dia que se deseja reconstruir. A Figura 2 apresenta um diagrama com as etapas do método e em seguida, é exposto detalhadamente cada etapa.

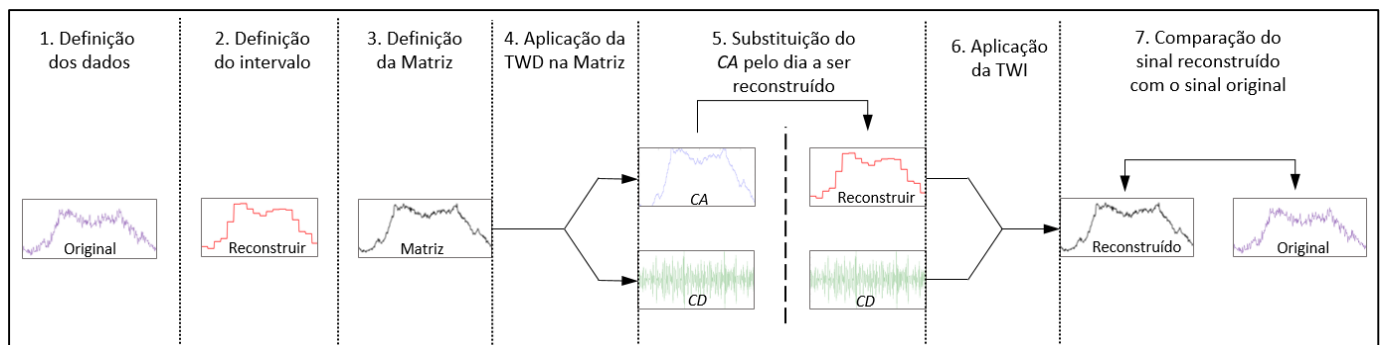

**Figura 2:** Método proposto para reconstrução do sinal

## 4.1. Definição dos dias a serem reconstruídos

Do banco de dados disponível, no total, 66 dias foram selecionados para a reconstrução seguindo o critério: três dias úteis típicos (terça, quarta e quinta-feira) de 1 semana por mês do período de março/2012 (mês da Matriz) até maio/2014. A escolha desses dias úteis típicos segue o método descrito por Ribeiro (2017), o qual considera o mês e os dias típicos onde não ocorreram erros no sensor e nenhum feriado que ocasionasse interferências no comportamento do tráfego. E, a escolha de uma semana por mês se deve ao fato de, na maioria dos meses, os dados sofreram interferência por falhas do sensor ou o comportamento do tráfego era influenciado por feriados ou férias. Os meses de outubro-2012, novembro-2012, janeiro-2013 e fevereiro-2013 foram descartados por falhas constantes dos sensores.

## 4.2. Preparação do intervalo

Em Estudos de Tráfego comumente os dados estão agregados em intervalos de 15 minutos, 1 hora ou, as vezes, 1 dia. A Transformada *Wavelet* Discreta tem limitação quanto a quantidade de níveis que pode ser aplicado e quantas informações resulta cada nível. Apesar dos dados de contagem geralmente estarem em intervalos de 15 minutos, devido ao *downsampling*, a taxa de amostragem da Transforma *Wavelet* Discreta reduz pela metade a cada nível de decomposição. Portanto, não é possível decompor um sinal de 5 minutos e reconstruir com dados agregados em 15 minutos, por exemplo. Como o banco de dados original tem intervalo ($t$) de 5 minutos, cada nível de decomposição ($n$) redimensiona o sinal em uma nova janela de intervalo $t \times 2^n$. A Tabela 2 apresenta a escala da janela para cada nível de decomposição da *Discrete Wavelet Transform*.

**Tabela 3:** Escala da janela em cada nível de decomposição

| Nível ($n$) | Janela de tempo ($5 \times 2^n$) | |
|---|---|---|
| 1 | 10 min | |
| 2 | 20 min | |
| 3 | 40 min | |
| 4 | 80 min | (1h20min) |
| 5 | 160 min | (2h40min) |
| 6 | 320 min | (5h20min) |
| 7 | 640 min | (10h40min) |
| 8 | 1,280 min | (21h20min) |
| 9 | 2,560 min | (1dia-18h40min) |
| 10 | 5,120 min | (3dias-13h20min) |
| 11 | 10,240 min | (7dias-02h40min) |

Nota-se na Tabela 3 que os coeficientes que mais se aproximam de uma hora, são os dos níveis três e quatro, 40 e 80 minutos respectivamente. Portanto, este artigo analisou 4 reconstituições, para os dados agregados em 10 minutos (1 nível de recomposição), em 20 minutos (2 níveis), 40 minutos (3 níveis) e por fim, para 80 minutos (4 níveis).

## 4.3. Determinação dos dias utilizados como dados desagregados – Matriz

Conforme exposto na Seção 2, para aplicar a Transformada *Wavelet* Inversa são necessários os componentes de Detalhe e Aproximação. Como o objetivo da pesquisa é reconstruir dados desagregados a partir de dados agregados, foi selecionado um conjunto de dias distintos aos dias que seriam reconstruídos para compor a Matriz e assim, obter o coeficiente de Detalhe. Apesar da Matriz ser composta por dias distintos, a mesma também foi obtida no sensor do km 51,9 da Rodovia Presidente Castelo Branco e também foi baseada na escolha dos três dias úteis típicos (terça, quarta e quinta-feira), porém apenas no mês de março/2012 (mês livre de feriados e eventos atípicos na região em que o sensor estava localizado). No total, a Matriz é composta por 13 dias indicados na Figura 3.

| **Março/2012** | | | | | | |
|---|---|---|---|---|---|---|
| D | S | T | Q | Q | S | S |
|  |  |  |  | 1 | 2 | 3 |
| 4 | 5 | 6 | 7 | 8 | 9 | 10 |
| 11 | 12 | 13 | 14 | 15 | 16 | 17 |
| 18 | 19 | 20 | 21 | 22 | 23 | 24 |
| 25 | 26 | 27 | 28 | 29 | 30 | 31 |

Legenda:
- Dias selecionados para Matriz
- Fim de semana

**Figura 3:** Dias selecionados para compor a Matriz

No entanto, ao utilizar os dados agregados para Matriz pode ocorrer perda de informações no coeficiente de Detalhe. Para verificar essa perda, os dados da Matriz foram utilizados em dois cenários distintos, conforme a seguir.

### 4.3.1. Cenário 1
No Cenário 1 foi mantido a média do fluxo de tráfego a cada cinco minutos dos dias selecionados. Nesse cenário, não há alteração de escala dos dados originais da Matriz, portanto não há nenhuma interferência na escala original em que os dados foram fornecidos (5 minutos) da Matriz.

### 4.3.2. Cenário 2
No Cenário 2 foi determinado a taxa de fluxo equivalente para 20 minutos. Foi verificado esse cenário para avaliar se uma Matriz com janela maior, porém com a mesma quantidade de dados do sinal original interferiria na reconstituição dos dados agregados. A Figura 4 apresenta os dois sinais discretos utilizados como Matriz no Cenário 1 e no Cenário 2, sendo em (a) o dia inteiro e (b) intervalo de 2 horas. Apesar dos dois cenários terem sido originados dos mesmos dados, o Cenário 1 mantém as características a cada 5 minutos e o Cenário 2 uma média no intervalo de 20 minutos.

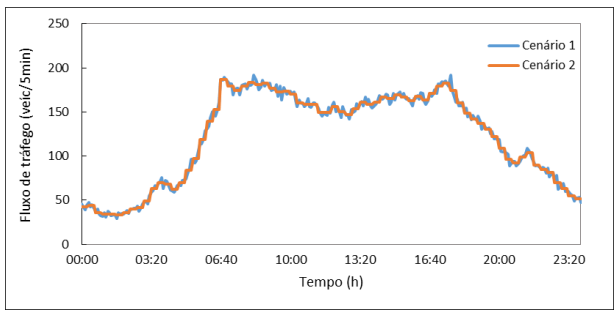 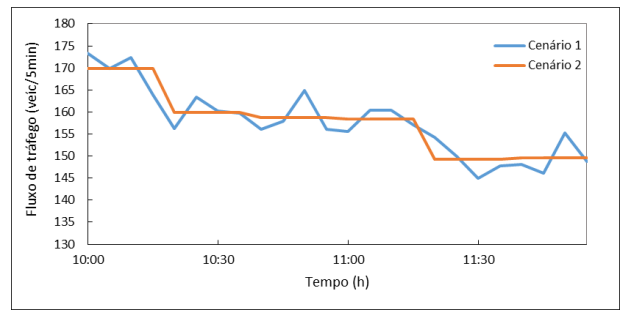

a) Sinal Completo da Matriz          b) Sinal da Matriz para o intervalo de duas horas

**Figura 4:** Sinal discreto da Matriz em função do tempo para os dos Cenários

### 4.4. Aplicação da TWD na Matriz
Após determinar a Matriz, foi aplicado a Transformada *Wavelet* Discreta de Haar em cada nível de decomposição pré-estabelecidos (1, 2, 3 e 4). Ressalta-se que a partir da Matriz é gerado apenas um coeficiente de Detalhe e de Aproximação para cada nível, ou seja, o mesmo coeficiente de Detalhe é utilizado para reconstrução de todos os dias.

### 4.5. Substituição do coeficiente de Aproximação pelos dias a serem reconstruídos
Ao aplicar a TWI nos coeficientes de Aproximação e Detalhe gerados pela TWD sem nenhuma alteração, o resultado será um sinal idêntico ao que foi decomposto. Portanto, se fosse aplicado a TWI nos coeficientes gerados pela TWD da Matriz, o resultado seria um sinal idêntico a Matriz. Como o intuito é reconstruir outro conjunto de sinais, foi retirado o coeficiente de Aproximação e substituído pelos dados agregados (10, 20, 40 e 80 minutos) dos dias que serão reconstruídos. Essa etapa é fundamental para o método, pois o coeficiente de Aproximação é que armazena as informações do comportamento padrão do dia. Com essa substituição é retirado o comportamento da Matriz e inserido o comportamento do dia a ser reconstruído.

### 4.6. Aplicação da TWI

Com o coeficiente de Detalhe gerado pela TWD aplicado na Matriz e com o coeficiente de Aproximação substituído pelos dados agregados do dia a ser reconstruído, foi aplicada a Transformada *Wavelet* Inversa de Haar (mesma utilizada na TWD). A quantidade de níveis executada pela TWI é correspondente a janela de tempo dos coeficientes, ou seja, para dados agregados em 10 minutos, 1 nível, para dados agregados em 20 minutos, 2 níveis e assim sucessivamente até 4 níveis.

### 4.7. Comparação dos dados reconstruídos com os dados reais

Independentemente da janela de agregação dos dados no coeficiente de Aproximação, a TWI foi aplicada na quantidade de níveis necessários para reconstrução dos dados em 5 minutos (sinal original). Ou seja, foi reconstruído os dados desagregados em 5 minutos a partir de dados agregados do mesmo período e do coeficiente de Detalhe da Matriz. Para verificar a eficiência do método, os resultados são avaliados de acordo com a correlação e o erro médio absoluto entre os dias reconstruídos e o sinal original, conforme é detalhe nos resultados na Seção 5, a seguir.

## 5. RESULTADOS
### 5.1. Cenário 1

A Figura 5 ilustra parte do sinal reconstruído (06:00 às 12:00) do primeiro dia reconstituído (13/03/2012) como exemplo e apresenta a reconstrução do sinal para quatro situações, dados agregados em: (a) 10; (b) 20; (c) 40 e; (d) 80 minutos com intervalo de 5 minutos. Ao utilizar a Transformada *Wavelet* Inversa no coeficiente de Detalhe da Matriz ao invés de um coeficiente de Detalhe do próprio sinal original ocorre uma deturpação quanto a escala em relação aos dados originais, pois a resposta do coeficiente de Detalhe está em energia e cada sinal tem suas próprias oscilações. Dessa forma, o sinal reconstruído manteve as informações dos dados agregados (coeficiente de Aproximação) mas em outra proporção. Para corrigir esse impasse, tanto o sinal original quanto o sinal reconstruído foram transformados em porcentagem dividindo o fluxo de tráfego de cada janela pelo Volume Total Diário do sinal, ou seja, cada 5 minutos do sinal original foi dividido pelo Volume Total Diário do sinal original e a cada janela de 5 minutos do sinal reconstruído foi dividido pelo Volume Total Diário do sinal reconstruído. Dessa forma, tanto o sinal original quanto o sinal reconstruído mantiveram suas características quando ajustada a proporção, o que permite realizar as análises necessárias de um em relação ao outro.

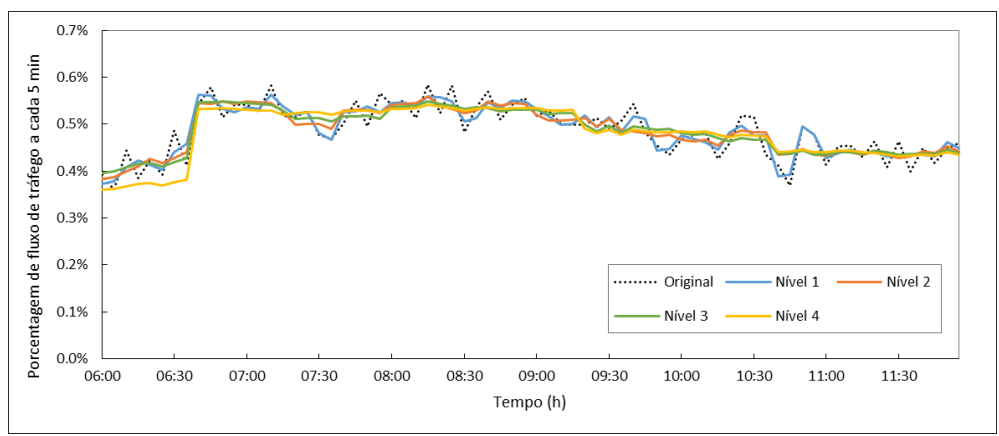

**Figura 5:** Intervalo de 6 horas do dia 13/03/2012 reconstruído em 4 níveis – Cenário 1

Na Figura 5, nota-se que conforme aumenta os níveis de decomposição/reconstrução, mais suave ficam os sinais reconstituídos mesmo com intervalo de 5 minutos. Isso pode ser explicado por dois motivos: (1) a Matriz é composta por uma média de todos os dias úteis típicos do mês, o que torna a mesma um sinal mais suavizado em relação a qualquer outro dia útil típico, mesmo mantendo o intervalo de agregação igual o banco de dados original (5 minutos); (2) apesar dos coeficientes de Detalhe armazenarem as informações de frequência da Matriz independentemente da quantidade de decomposição, o Coeficiente de Aproximação foi substituído pelo sinal agregado em até 1 hora e 20 minutos (4 níveis de decomposição) o que também mitiga o sinal e perde informações para reconstrução.

Para avaliar o desempenho da reconstrução utilizada no método proposto, foi empregada a correlação e o erro absoluto médio entre a porcentagem do sinal original e dos sinais reconstruídos. A Tabela 4 apresenta o resumo dessas análises para cada nível de decomposição, a Figura 6 apresenta a correlação entre o sinal original e o sinal reconstruído para os 4 níveis enquanto a Figura 7 apresenta o erro absoluto médio.

**Tabela 4:** Correlação e Erro Absoluto Médio entre a o sinal original e reconstruído – Cenário 1

| Resolução dos Dados | Níveis | Correlação | | | | Erro Absoluto Médio | | | |
|---|---|---|---|---|---|---|---|---|---|
| | | Média | Mediana | Máxima | Mínima | Média | Mediana | Máxima | Mínima |
| 10 min | 1 | 0,983 | 0,983 | 0,988 | 0,974 | 7,15% | 7,11% | 11,68% | 5,91% |
| 20 min | 2 | 0,974 | 0,976 | 0,982 | 0,956 | 8,77% | 8,71% | 12,87% | 7,45% |
| 40 min | 3 | 0,967 | 0,970 | 0,979 | 0,929 | 9,85% | 9,68% | 14,09% | 8,60% |
| 80 min | 4 | 0,959 | 0,963 | 0,973 | 0,912 | 11,26% | 11,09% | 15,44% | 10,02% |

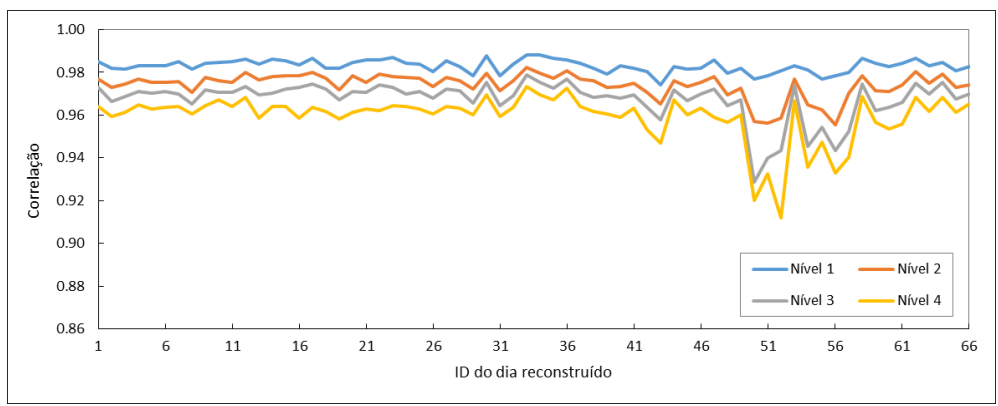

**Figura 6:** Correlação entre o sinal original e o sinal reconstruído nos quatro níveis de reconstrução – Cenário 1

Nota-se alta correlação entre o sinal original e o sinal reconstruído para os quatro níveis de reconstrução, sendo a mínima de 0,912 para os dados reconstruídos em quatro níveis (dados agregados em 80 minutos) no dia 52 e máxima de 0,988 para os dados reconstruídos em um nível (dados agregados em 10 minutos) nos dias 30, 33 e 34. Observa-se que o sinal reconstruído a partir de menos dados agregados obtiveram uma maior correlação com o sinal original, o que corrobora com a influência do coeficiente de Aproximação para a reconstrução com a TWI. A média e a mediana da correlação para os quatro níveis foram similares, o que indica uma distribuição simétrica, todos os dias reconstruídos foram bem correlacionados aos dados originais.

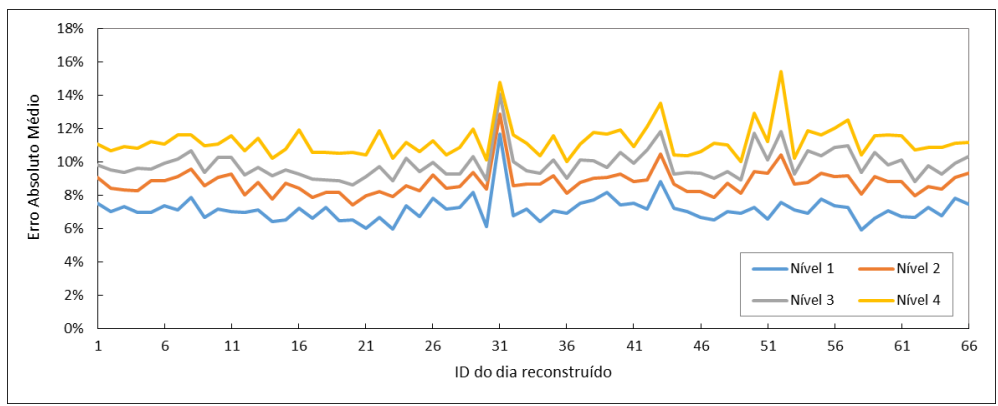

**Figura 7:** Erro Absoluto Médio entre o sinal original e o sinal reconstruído nos quatro níveis de reconstrução – Cenário1

Ao avaliar o Erro Absoluto Médio também foi possível verificar o desempenho da reconstrução dos dados. Nota-se que os dados agregados em 10 minutos tiveram um melhor desempenho, com o menor erro médio absoluto de 5,91% no dia 58. Por outro lado, os dados agregados em 80 minutos tiveram erro médio absoluto mínimo de 10,02% no dia 49 e erro médio absoluto máximo de 15,44% no dia 52, apesar disso, como visto anteriormente, a correlação foi similar ao nível 1. Entende-se que o sinal reconstruído manteve uma relação de similaridade ao sinal original, mas como no nível 4 o mesmo está suavizado e o sinal original não, o erro médio absoluto foi maior. Tais resultados podem ser equiparados à estudos de previsão de fluxo de tráfego (CASTRO-NETO *et al.,* 2009; CORIC et al. 2012, LAM *et al.,* 2006; LIM, 2001).

**5.2. Cenário 2**

A Figura 8 ilustra o primeiro dia reconstituído (13/03/2012) como exemplo e apresenta a reconstrução do sinal para quatro níveis, além do sinal original.

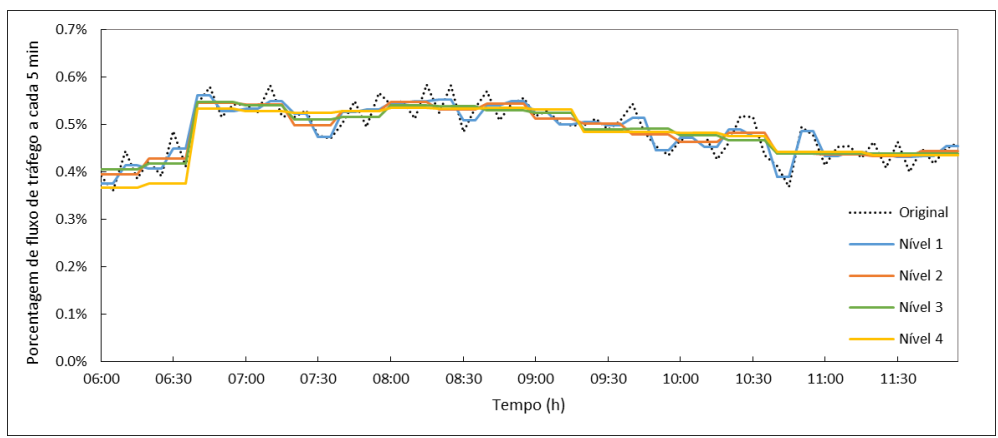

**Figura 8:** Intervalo de 6 horas do dia 13/03/2012 reconstruído em 4 níveis – Cenário 2

Para o Cenário 2, o sinal da Matriz foi constituído a partir da taxa de fluxo equivalente para 20 minutos. Dessa forma, os níveis 2, 3 e 4 (20, 40 e 80 minutos, respectivamente) ao serem reconstruídos mantiveram o valor a cada 5 minutos dentro da janela de 20 minutos. Isso por que o coeficiente de Detalhe para o nível 1 e 2 na decomposição da Matriz resultou em zero. A TWD decompõe aos pares e vizinhos, como nessa situação os valores eram os mesmos a cada quatro dados, não há diferença de frequência, sendo assim, zero (Figura 9). Por outro lado, para a reconstrução dos dados agregados em 10 minutos, devido a janela do Coeficiente de Aproximação, o sinal reconstruído manteve o intervalo de 10 minutos apesar do coeficiente de Detalhe (20 minutos).

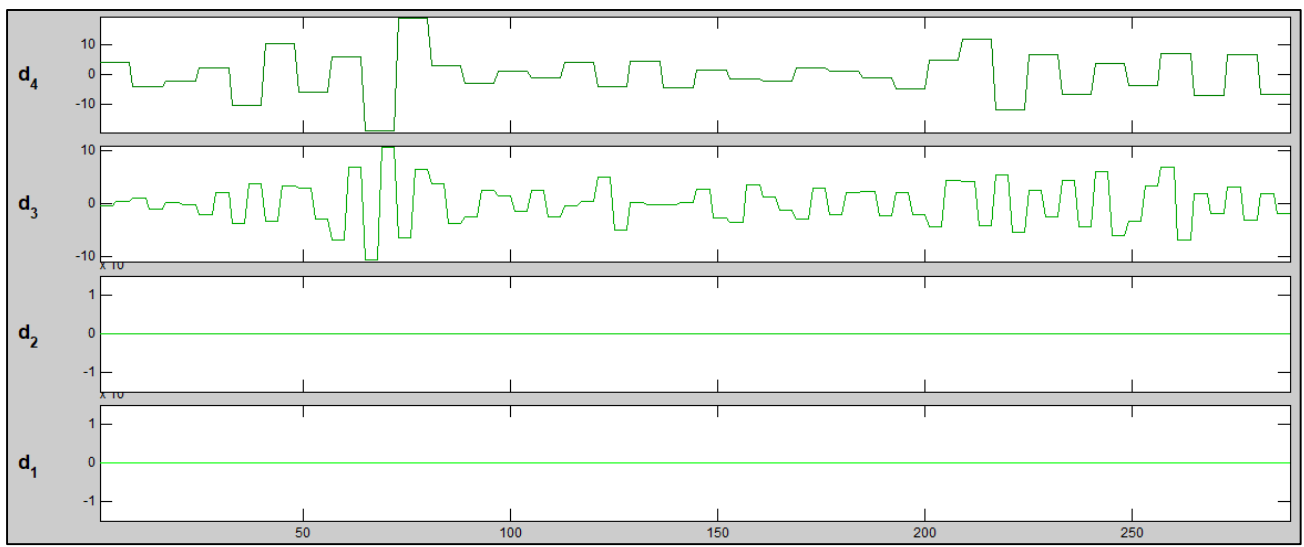

**Figura 9:** Coeficientes de Detalhe 1, 2, 3 e 4 para a Matriz do Cenário 2

Na avaliação do desempenho da reconstrução do sinal pelo método adotado, também foi empregada a correlação e o erro absoluto médio entre o sinal original (valor original a cada 5 minutos) e os sinais reconstruídos. A Tabela 5 apresenta o resumo dessas análises.

**Tabela 5:** Correlação e Erro Absoluto Médio entre a o sinal original e reconstruído – Cenário 2

| Resolução dos Dados | Níveis | Correlação | | | | Erro Absoluto Médio | | | |
|---|---|---|---|---|---|---|---|---|---|
| | | Média | Mediana | Máxima | Mínima | Média | Mediana | Máxima | Mínima |
| 10 min | 1 | 0.983 | 0.983 | 0.988 | 0.974 | 7.06% | 7.01% | 11.49% | 5.86% |
| 20 min | 2 | 0.974 | 0.975 | 0.982 | 0.954 | 8.85% | 8.83% | 13.08% | 7.69% |
| 40 min | 3 | 0.967 | 0.970 | 0.978 | 0.929 | 9.93% | 9.72% | 14.17% | 8.72% |
| 80 min | 4 | 0.959 | 0.962 | 0.973 | 0.912 | 11.31% | 11.16% | 15.52% | 10.01% |

Assim como o Cenário 1, o Cenário 2 também apresentou alta correlação e baixo percentual de erro absoluto médio. A correlação média dos dois cenários foi igual para todos os níveis e a mediana, máxima e mínima sofreu uma variação de ± 0,002 em alguns níveis. Por outro lado, o erro absoluto médio variou em ± 0,09% na média, ± 0,12 na mediana, ± 0,21 no valor máximo e ± 0,24% no valor mínimo. Esses resultados indicam novamente a influência na agregação dos dados, porém, no Cenário 2 influenciando negativamente os coeficientes de Detalhe da Matriz ao eliminar diferenças de frequências existente nos sinais da Matriz e dias reconstruídos a cada 5 minutos.

## 6. CONCLUSÕES E RECOMENDAÇÕES

Este artigo propôs um método de reconstrução de dados históricos de fluxo de tráfego desagregados (5 minutos) a partir de dados agregados (80 minutos) a partir da Transformada *Wavelet* Inversa. Para tal, foi proposto como coeficiente de Aproximação os dados agregados do dia a ser reconstruído e como coeficiente de Detalhe, o componente de Detalhe gerado da Transformada *Wavelet* Discreta empregada em uma Matriz constituída de dias distintos aos dias a serem reconstruídos.

Apesar da Transformada *Wavelet* Inversa ser capaz de reconstruir um sinal decomposto de forma integra, sem variação, o método proposto neste trabalho não foi capaz de fazer o mesmo. Isso porque ao empregar médias para formação da Matriz e agregação do sinal a ser reconstituído, ocorreu suavização de frequências típicas do comportamento do fluxo de tráfego, indicando assim uma interferência negativa das médias na reconstrução dos dados, tanto no coeficiente de Aproximação quanto no de Detalhe.

No entanto, a correlação entre os dias reconstituídos e o sinal foi alta, assim como o erro absoluto médio foi baixo, indicando a eficiência do método. Apesar dos resultados obtidos terem sido satisfatório, e comparáveis a estudos de estimativa de tráfego (CASTRO-NETO *et al.,* 2009; CORIC *et al.* 2012*,* LAM *et al.,* 2006; LIM*,* 2001), o intuito deste trabalho é oferecer ferramentas para construção de banco de dados que seja útil para as diversas áreas da Engenharia de Transportes que avaliam dados em nível de representatividade macroscópica.

Para pesquisas futuras, recomenda-se a aplicação do método proposto para construção de bancos de dados de rodovias em que haja pouca ou nenhuma informação sobre o fluxo de tráfego. Além disto, estudos futuros podem avaliar Matrizes em intervalos de tempo menores para aumentar a correlação do sinal original e o sinal reconstruído. A extensão do trabalho aqui apresentado pode abordar a reconstrução de outros parâmetros de tráfego, como por exemplo, a velocidade média.